\documentclass[11pt,twoside]{article}
\usepackage{asp2004}
\usepackage{psfig}
\usepackage{epsf}
\usepackage{graphics}
\usepackage{lscape}
\def\edcomment#1{\iffalse\marginpar{\raggedright\sl#1\/}\else\relax\fi}
\reversemarginpar

\begin{document}
\title{Where Are We Now, and Where Are We Going?  Perspectives on Cosmic
Abundances}

\author{Timothy C. Beers}
\affil{Department of Physics and Astronomy, Michigan State University, and JINA: Joint Institute for Nuclear
Astrophysics, E. Lansing, MI  48824}

\begin{abstract}
I was asked at the end of this (excellent!) conference to provide a few comments on the
present status of the field of cosmic abundances, which I summarize (and extend
upon) in this contribution.  A discussion of the present ``state of play'' is
given, along with some suggested directions for the future.  
\end{abstract}

\thispagestyle{plain}

\section{Introduction}

Over the past decade we have witnessed an explosion of new information
pertaining to the topic of cosmic abundances, spurred primarily by
high-resolution spectroscopic investigations of individual stars in the halo and
thick disk of the Milky Way and its satellites, as well as in clouds of gas
observed in the directions toward distant quasars. These two sets of probes,
which represent vastly different scales in distance, broadly overlap with one
another on the scales of the look-back time they explore. None of this recent
progress would have been possible without reasonable access of the abundance
community to 8m-10m class telescopes equipped with modern, high-throughput
spectrographs. This conference has provided a nice ``snapshot'' of accomplishments to
date, as well as an opportunity for practitioners of the art of abundance
determination and analysis to consider the directions that we might hope to
pursue in the future. Below I offer a few perspectives on these issues,
concentrating on stars of very low metallicity.

\section{Where Are We Now?}
\bigskip

Before looking too far forward it is good to glance over our shoulders, and
take note of how far we have come. 

\subsection{Surveys for Metal-Poor Stars in the Past and Present}

There are two primary surveys for low-metallicity stars that have been
carried out over the course of the past two decades. The HK survey of
Beers and colleagues (formerly known as the ``Preston-Shectman survey''; Beers,
Preston, \& Shectman 1992; Beers 1999) has identified over 1000 Very Metal-Poor
(VMP) stars in the halo and thick disk of the Galaxy, with [Fe/H] $< -2.0$, on the
order of 100 Extremely Metal-Poor (EMP) stars, with [Fe/H] $< -3.0$, and a
handful of stars with [Fe/H] $\approx -4.0$. There are no Ultra Metal-Poor (UMP)
stars (yet) found in the HK survey, with metallicities significantly below [Fe/H] =
$-4.0$. These numbers only apply to the HK-I survey, which was based on visual
selection of low-metallicity candidates from the prism plates. Rhee and collaborators (see
contribution from Rhee in this volume) are presently extending the HK survey,
based on digitized scans of the HK prism plates carried out with the APM in
Cambridge. HK-II, as this effort is known, is able to identify large numbers of
relatively bright ($B \le 15.5$) metal-poor giants, many of which were originally
overlooked in HK-I due to an unavoidable temperature-related bias in the visual
selection of candidates. 

The second large survey effort is that of Christlieb and colleagues
(Christlieb 2003), known as the Hamburg/ESO Survey (HES). Based on digitized
scans of southern-hemisphere prism plates, this survey has (to date) yielded a
similar number of [Fe/H] $< -2.0$ stars as the HK survey, but, significantly,
has at least doubled the numbers of stars known with [Fe/H] $< -3.0$. Of
particular importance, the ``reach'' of metal deficiency has now been extended
to include Hyper Metal-Poor (HMP) stars ([Fe/H] $ < -5.0$), as a result of the
discovery of the most iron-poor star yet known, HE~0107-5240, with [Fe/H] =
$-5.3$ (Christlieb et al. 2002, 2004a). Continued medium-resolution follow-up
spectroscopy of the several thousand remaining low-metallicity candidates in the
HES may well reveal additional examples of HMP stars, or even (if they exist), Mega
Metal-Poor (MMP) stars, with [Fe/H] $< -6.0$.

\subsection{Abundance Trends and Scatter}

One of the main reasons the surveys for metal-poor stars were carried out
in the first place was to enable high-resolution spectroscopic studies of
individual elemental abundances for significant numbers of VMP stars.  In the
past several years we have begun to reap this harvest of information on the
nature of the chemically most primitive stars in the Galaxy; numerous
contributions in this conference have reported on these results.  In the
published literature, the interested reader is directed to the series of papers
on the ``First Stars'' Large Program conducted with VLT/UVES (e.g., Cayrel et al.
2004), as well as the ``0Z'' program being conducted with the Keck telescope
(e.g., Cohen et al. 2004). 

Two of the important results that are emerging from these high-resolution
spectroscopic analyses are the identification of well-defined trends of elemental
abundances with [Fe/H] (or [Mg/H]), as well as quantitative determination of the
cosmic scatter of these abundances about the general trends. Although the
character of the trends (at least for elements in common) have not changed much
compared to what was known a decade ago (see, e.g., McWilliam et al. 1995;
Norris, Ryan, \& Beers 1996), there has been a rather dramatic change in the
reported scatter about these trends. In particular, the new studies have shown
that much of the previously claimed scatter for alpha- and iron-peak elements at
low metallicities was spurious, and due to limitations in the quality of the
available data, or the addition of ``observer scatter'' arising from the
practice of combining small individual samples from multiple sources (and hence
importing differences in analyses). The new (much higher-quality) data have
revealed that the observed scatter for many elements in VMP stars is near the
level of the observational error, i.e., consistent with zero. For example,
the reported scatter in [Cr/Fe] by Cayrel et al. (2004), based on a large sample
of VMP giants, is only 0.07 dex. Arnone et al. (2004) report, based on
high-resolution spectroscopy of a sample of VMP main-sequence turnoff stars, a
scatter in [Mg/Fe] of only 0.06 dex. The implications of these results for models of
supernovae yields, early star and galaxy formation, and galactic chemical
evolution, will be profound.

\subsection {Rosetta Stars}

The literature is rapidly becoming populated with examples of individual stars
that provide unique windows on the nature of element production in the early
universe. These include r-process-enhanced EMP stars (e.g., CS~22892-052,
Sneden et al. 2003; CS~31082-001, Hill et al. 2002), VMP stars with large
over-abundances of s-process elements (e.g., CS~29497-030, Sivarani et al.
2004), and EMP stars with large enhancements of light elements such as C,N,O,
and the alpha elements (e.g., CS~29498-043, Aoki et al. 2004), to name a
few. Such stars are important for several reasons. First, their very existence
is guiding the development of new models of the astrophysical scenarios
responsible for the production of these characteristic signatures. Secondly, now
that these objects have been detected, new surveys can (and are) being carried
out to find numerous additional examples of these phenomena.  

\section{Where Are We Going?}

Looking ahead, we can anticipate even more exciting times ahead for members of
the abundance community.

\subsection {Surveys for Metal-Poor Stars in the (Near) Future}

Rapid progress can only be made with an expansion of the present efforts to mine
the Galaxy for ever increasing numbers of VMP stars. As examples of where things
are headed, I mention two surveys, with which I am most familiar, that are
presently underway or will begin in the next year.  

Christlieb et al. (2004b) describe an ongoing high-resolution spectroscopic
follow-up of VMP giants selected from the HK and HES surveys directed at the
identification of (rare) additional examples of stars with large over-abundances
of r-process elements. The Hamburg/ESO R-process-Enhanced Star (HERES) survey
has already obtained ``snapshot'' (15-20 minute integrations) spectroscopy of
over 370 VMP giants with VLT/UVES. In addition to r-process-enhanced stars (at
least 7-10 new such stars have already been identified), the HERES survey is
providing numerous new examples of s-process-enhanced stars, carbon-enhanced
stars, and stars that exhibit signatures of both r- and s-processing (r/s stars).
Of particular importance, the data is of sufficient quality to determine
elemental abundances (with somewhat larger errors than are now obtainable with
higher signal-to-noise data) for on the order of 15-20 elements per star. The
HERES targets were selected based exclusively upon their metal
deficiency (as revealed from medium-resolution follow-up of prism-survey
candidates). Hence, HERES will provide the first {\it large} sample of VMP stars,
analyzed in a homogeneous fashion, from which one will be able to obtain
estimates of the frequency of occurrence of many of the Rosetta stars mentioned
above, as well as provide the raw material upon which self-consistent models of
supernova yields and galactic chemical evolution can be constructed.  

Plans are presently being made, and tests of target selection and analysis
techniques are being completed, for the next great expansion of our
observational database of interesting stars in the Galaxy. This survey effort,
known as SEGUE: Sloan Extension for Galactic Understanding and Evolution, will,
as the acronym suggests, be employing the same telescope (the ARC 2.5m
telescope on Apache Point, New Mexico), imager, spectrographs, and reduction
pipeline as the original SDSS to dramatically enhance our knowledge of the
stellar populations of the Milky Way. SEGUE, presently scheduled to start in
July 2005, and finish by July 2008, will obtain some 3000 square degrees
of calibrated $ugriz$ photometry at lower Galactic latitudes than the SDSS main
survey, so as to better constrain the important transition from the disk
population(s) to the halo. Most importantly, SEGUE will obtain medium-resolution
($R = 1800$) spectroscopy for 250,000 stars in the magnitude range $14 \le g \le
21$, in 200 directions covering the sky available from Apache Point. SEGUE
targets have been chosen to explore the Galaxy at distances from 0.5 -- 100 kpc
from the Sun. It is no exaggeration to claim that SEGUE will completely
revolutionize our understanding of the structure and stellar populations of the
Milky Way.

One of the primary SEGUE categories is obtained from photometric pre-selection
of likely VMP stars.  Tests carried out to date indicate that it is reasonable
to expect that SEGUE will yield a sample of some 50,000 stars with [Fe/H] $<
-2.0$, a factor of {\it 20 to 25 times} the present number of such stars known from the
summed efforts of astronomers to date (essentially the HK and HES surveys). 
Although the majority of these VMP stars will be too faint for easy
high-resolution spectroscopic follow-up with 8m-10m telescopes (but tractable
with 30m-100m telescopes !), there will be at least several thousand that are
sufficiently bright.  Plans are presently being made to undertake HERES-like
snapshot spectroscopy with the Hobby-Eberly, Subaru, and (perhaps) the Keck
telescopes, followed by high S/N studies of the most interesting objects that
are revealed.  The next three years will certainly be exciting ones for stellar
spectroscopists !
 
\subsection{Analysis Techniques}

As should be obvious from the preceeding discussion, samples of stars with
available high-resolution spectroscopy will soon completely overwhelm
``traditional'' approaches to data reduction and analysis. Hence, as part of the
HERES development, we have implemented completely automated analysis methods for
obtaining elemental abundance estimates (see Barklem et al. 2004). It should be
feasible to implement such techniques via a web interface, so that they can be
carried out as (pipeline reduced) high-resolution data roll off the telescope.
The era of ``industrial'' abundance measurements is nearly at hand.

\subsection{Access to Results of Abundance Analyses}

Mass production of stellar elemental abundances will be of little use if they
cannot be cast into a form that the community of astronomers, and interested
parties outside of astronomy, can employ them in their own work. A familiar
frustration to those in the abundance game is the difficulty (often,
impossibility) of reliably combining high-resolution abundances obtained by
colleagues using different spectrographs, different spectral resolutions, different S/N
targets, different input physical parameter estimates, different atmospheric models,
different line lists, different solar reference abundances, different atomic
parameters, etc., into large homogeneous samples for further analysis. 

Fortunately, it is now feasible (although quite a bit of work initially) to set
up an ``abundance analysis clearinghouse'' that would be capable of taking input
reduced high-resolution spectra, smooth them to a common resolution (or set of
resolutions), and apply a completely automated, objective, and self-consistent
analysis. The advantages of such an approach are obvious, as the elemental
abundances that result will be on a common system that does not suffer from the
``observer scatter'' discussed above, and can be readily and confidently used by
specialists and non-specialists alike. This might be a task best implemented
under the auspices of the National Virtual Observatories now in existence, or
being set up worldwide. Moreover, as analysis techniques, atomic data, model
atmospheres, etc., improve in the future, the database of spectroscopy can be
quickly re-analyzed to take full advantage of this progress, something which is
rarely done at present. As the database of spectroscopy grows, it will also
provide a valuable resource for the design of new surveys, as one will be able
to easily check if suitable data for stars of interest already exist,
or whether there is a need to improve the data with additional observing.

As a first application, we soon hope to implement such an approach making use of
data that is now available from various archives in the public domain.

\subsection{Collaborations with Nuclear and Atomic Physicists}

In the past, our colleagues in the related sub-fields of nuclear and atomic
physics have often been referred to as {\it providers} of (some of) the
fundamental models and data required for successful elemental abundance analyses
and astrophysical interpretations by astronomers, the {\it consumers} of this
information. I believe that the time has come to break down some of the present
barriers (when is the last time {\it you} have attended an APS meeting, for
example), and seek out opportunities (which exist at present, and which will
expand in the near future) for the cosmic abundance community to link up
more closely with the communities that our nuclear and atomic physics colleagues
associate themselves with. 

I have already begun to assimilate. I am a co-PI in a recently funded NSF
Physics Frontier Center, JINA: Joint Institute for Nuclear Astrophysics, a
collaboration between the University of Notre Dame, Michigan State University,
and the University of Chicago. We have, over the course of the past year,
accreted a long list of universities, national labs, and other institutions
worldwide, as associate members of JINA. We welcome additional members. For more
information, please visit the JINA website: {\bf http://www.jinaweb.org/}.
JINA sponsors, or co-sponsors, three to five meetings per year related to cosmic
abundances. The primary mission of JINA is to develop and expand the community
of physicists and astronomers with interest in nuclear astrophysics, and to
attack complex problems that demand expertise from a broad range of specialists;
one example is the origin of the elements. It is safe to say that most, if not
all, of the attendees at this conference have a vested interest in this topic.
JINA maintains a ``virtual journal'' that assembles articles from the prominent
journals of astronomy and physics that pertain directly to nuclear astrophysics.
You are invited to visit the JINA website and take advantage of this resource.

Although it may not be widely known to astronomers, the nuclear physics
community has decided that the bulk of their effort over the next several decades
will be concentrated in the construction and use of a proposed new DOE-funded
facility capable of obtaining nuclear data for isotopes involved in the
production of the elements. The presently considered project, RIA: Rare Isotope
Accelerator, will be capable of measuring the fundamental properties of the
isotopes (e.g., masses, reaction cross sections, beta-decay rates, etc.) that
are created by the element-producing stars of interest to astronomers. It is
imperative, in my view, that astronomers in the cosmic abundance community
involve themselves in this effort (see {\bf http://www.nscl.msu.edu/ria/}). 

One final thought. Physics Frontier Centers presently represent a significant
fraction ($\sim $ 10\%) of the budget of the NSF physics division. Today, there
are 10 Physics Frontier Centers located around the country. Over the course of
the next five years, the physics division at NSF hopes to raise this number to
on the order of 20-25, depending on budget realities. To my knowledge, no such
centers are supported (solely) by the NSF astronomy division. I believe that, in
the near future, it is of particular importance to encourage the development of
Astronomy Frontier Centers, which could unite substantial numbers of faculty
investigators, students, and postdocs, to work together on the issues of central
importance to our field.  

\acknowledgements 

I would like to acknowledge partial funding, which enabled my attendance at this
meeting, from grant AST 04-06784, as well as from grant PHY 02-16783: Physics
Frontiers Center/Joint Institute for Nuclear Astrophysics (JINA), awarded by the
U.S. National Science Foundation.

\end{document}